\documentclass[preprint,12pt,groupaddress]{elsarticle}

\usepackage{booktabs}
\usepackage{color}
\usepackage{lipsum}
\usepackage{float}
\usepackage[titletoc,toc,title]{appendix}
\usepackage{amssymb}
\usepackage{epsfig}
\usepackage{amsmath}
\usepackage{times}
\usepackage{subfigure}
\usepackage{graphicx}
\usepackage{booktabs}
\usepackage{colortbl}
\usepackage{natbib}

\begin{document}

\title{Supply based on demand dynamical model}

\author{Asaf Levi}
\author{Juan Sabuco}
\author{Miguel A. F. Sanju\'an}

\address{Nonlinear Dynamics, Chaos and Complex Systems Group.\\Departamento de F\'isica, Universidad Rey Juan Carlos, Tulip\'an s/n, 28933 M\'ostoles, Madrid, Spain}

\date{\today}

\begin{abstract}
We propose and analyze numerically a simple dynamical model that describes the firm behaviors under uncertainty of demand forecast. Iterating this simple model and varying some parameters values we observe a wide variety of market dynamics such as equilibria, periodic and chaotic behaviors. Interestingly the model is also able to reproduce market collapses.\end{abstract}

\maketitle

\section{Introduction}

Firms need to make the decision of how many goods to supply before they even know how many goods the market will demand in the next sales season. This problem is known as {\em uncertainty of demand forecast} and it has been widely studied in economics and supply chain management \cite{Uncertenty and Production Planning}. Being successful in predicting the future demand might be crucial for the survival of any productive company in a competitive market. The standard microeconomics models of the firm assume perfect information, implying that the firm knows exactly the shape of the demand curve. Furthermore, these models assume static and independent demand and supply curves, so that the decisions made by the firm do not have any effect on the shape nor the slope of the supply and demand curves. In this position the firm is only maximizing profits and its actions have no influence on the global dynamics of the market. So that in the long run the system settles down in equilibrium. Since the publication of the classic paper of George A. Akerlof \cite{lemons}, new models have been proposed. For example, pricing models of the monopoly under uncertainty of demand, considering the demand as a stochastic function \cite{ Baron, Dana, Raviv}, focusing on the optimal price for the firm and less on the market dynamics. Complexity economics in contrast, focuses on the emerging market dynamics created by economic agents when they react to patterns created by their own interactions during the time they interact \cite{Brian Arthur, Farmer, Tesfatsion}. This approach focuses on the connectivity and the interdependences between economic agents and how they organize and interact to achieve their economic end. Modeling the economy in this way opens up a new world of possibilities, where equilibrium is one possible dynamics among many others that can emerge from these interactions. In recent years, policy making have adopted Dynamic Stochastic General Equilibrium models (DSGE), to better predict and even control the economy at the macro level \cite{dsge, dsgeW}. These models are built from three main blocks where each one is a representation of some economic agent or a group of agents. The demand block represents the consumption of households, firms and even the government. The supply block represents the productive agents of the economy and the policy block represents financial institutions like central banks \cite{introDsge}. These kind of models add to the general equilibrium models some simple dynamical interaction between the economical agents in addition to some stochastic external shocks. In supply chain management when the firm has data sets it almost always uses statistical methods like time series analysis or linear regression \cite{Chase} to estimate the future demand. We find the combination of the last three frameworks interesting in nonaggregable models at the micro level of the economy.

In this work, we will think about the market as a dynamical place where one firm is a price maker while it has limited information about the demand. We focus on firms whose commercial activity involves producing or buying some stocks of a certain good with the purpose of selling them to obtain profits. These firms, mainly small, medium or entrepreneurs do not spend much resources in demand forecasting, they rely mainly on their buyers expectations among limited data sets of past sales, for example small stores, retailers or small factories that their main revenue comes from certain holiday tradition. For simplicity, we will call suppliers, to all the agents that belong to this group.

We will study numerically a dynamical model that is built similarly to a DSGE model without the stochastic terms, focusing  on the micro level of the economy. This model is highly inspired by the classic cobweb model~\cite{Cobweb theorem} with the difference that the supplier decision of how many goods to produce is sensitive to the quantity demanded instead of the market price of the good. The following two key concepts in the supply organization are captured in the model. First, as in the nonlinear version of the cobweb model, we present one possible dynamical procedure based on suppliers expectations \cite{Hommes, Nerlove, Chiarella, Artstein}, that can lead to market equilibrium and to chaos as well. The second idea embodied in this work, is that for a given quantity of supply the supplier fix some price that generates a demand feedback from the market. This information is needed to compute the quantity of supply in the next time step. As in real markets the supplier reacts to these demand feedbacks, what creates a rich price-quantity dynamics. Additionally, we will show that in some cases the supplier may push the market towards an equilibrium motivated by his selfish interests, sell all the stock, as Adam Smith once wrote: {\em ``It is not from the benevolence of the butcher, the brewer, or the baker that we expect our dinner, but from their regard to their own self-interest..."}. But in other cases the supplier may produce irregular dynamics that may lead to market collapse. We have found that the price elasticity of demand (PED) and the gross margin can play an important role in the stabilization of prices in the same way they can make the market crash.

The structure of the paper is as follows. Section 2 is devoted to the description of the supply based on demand model. Two types of suppliers and their behaviors are described in Section 3. In Section 4 we explore the dynamics of the model for several parameters. The global dynamics and results are described in section 5. In Section 6 we emphasize the idea that the final bifurcation means - market collapse. We describe the influence of the price elasticity of demand (PED) on the global dynamics in Section 7. Finally, some conclusions are drawn in Section 8.

\section{Description of the supply based on demand model}

We consider a supply and demand model of the form,
\begin{eqnarray}\label{model}
&D_{n+1} = a - bP_{n+1},\label{D}\\
&S_{n+1} = D^{Exp}_{n+1},\label{S}\\
&P_{n+1} =\frac{ATC}{1-M},\label{P}\
\end{eqnarray}
where the quantities demanded and supplied, $D_{n+1}$, and $S_{n+1}$, and the price, $P_{n+1}$ are assumed to be discrete functions of time. The parameters $a$ and $b$ are positive constants $a,b \geq 0$ and  $D^{Exp}_{n+1}$, is the expected demand. The parameter $M$, is the gross margin added by the supplier to obtain profits, where $0 \leq M < 1$ and ATC is the average total cost function of the good, that we will explain in details later on.

The quantity demanded in the market depends mainly on the price of a given good. The price of the good in contrast, depends heavily on the average total cost function, which is directly linked to the quantity of supply. When the supplier decides how many goods to produce, he always estimates in some way the future quantity of demand, $D^{Exp}_{n+1}$. The problem is, that the supplier makes the decision of what quantity to supply, $S_{n+1}$, before he knows the reaction of the market to the price that he fixes. In this model we assume, the supplier does not know anything about the demand function. The only available information he has, is the quantity demanded at the price in which he sold his products in the last sales seasons. We assume an ordinary good market in which, when the price increases, the consumption of the good decreases and vice versa. For simplicity, we assume a linear demand curve with negative slope as shown in Eq. (1). Before we proceed, we introduce two more mechanistic assumptions, that describe how the supplier operates in the market.\

{\bf Assumption 1}\\
{\em The supplier is the only one who sets and adjusts the price in light of circumstances.}

In this model the supplier is the only one who sets and adjusts the price. Notice that after the supplier launches the goods into the market, no changes can be done in the quantity supplied nor the price. The price structure is given by the ATC function and the gross margin as shown in Eq. (3). Both building blocks are known and controlled by the supplier.

\begin{figure}
\begin{center}
\includegraphics[width=0.8 \textwidth]{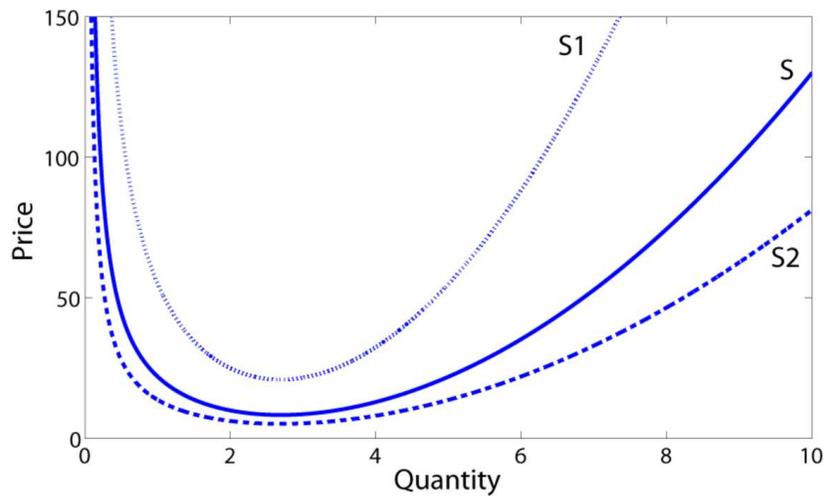}
\caption{\textbf{Parabolic price function.} We have used the following function                               $P = \frac{1}{{1-M}}\cdot({F_{c}}{Q}+v-vQ+Q^2)$, to relate the price of the good with the quantity supplied, where $P$ is the price of the good, $Q$ is the quantity, ${F_{c}}$ is the fix cost of production and $v$ is the variable cost of production. Notice that the ATC function inside the brackets, determines the parabolic shape of the price function. The parameters are fixed as: ${F_{c}} = 10$ and $v = 4$. The supply curves $S$ as solid line, $S1$ as dot line and $S2$  as dash-dot line, correspond to the gross margin $M = 0.5$, $M = 0.8$, $M = 0.2$ respectively. When the supplier increases the gross margin $M$, the price of a given quantity of goods increases as well.}\label{ATC}
\end{center}
\end{figure}

After estimating the demand for the next period, the supplier begins the production phase. He introduces his estimations in the ATC function to obtain his average total costs of production. We assume in this model that the average total costs is computed adding the fix costs to the variable cost per unit of good, divided by the total amount of goods produced. However, there are many possible ways to describe an ATC function. For instance, in many industries the prices lists shown to the buyers are organized in a ``piecewise function" fashion, where the price of the good is well established for every subset of quantities the buyer is willing to buy. But here, to stay faithful to the classical cost theory, we have chosen a typical continuous cubic total cost function, that gives rise a parabolic ATC function that depends also on the quantity of production $Q$ \cite{Holt Simon} as shown in Fig.~\ref{ATC}. In some markets the average total cost decreases as the supplier increases the amount of goods he produces or the amount of goods he is willing to buy, until reaching some critical point. After crossing this point every additional product produced or bought increments the average total cost. The parabolic shape of the ATC function as shown in Fig.~\ref{ATC} captures this idea. In the classic supply and demand model is taken for granted the linear positive slope shape of the supply curve what guarantees convergence towards an static equilibrium. In our case the supply curve is nonlinear, what produces more complex dynamics. The quantity of production $Q$ is the same as the quantity of supply, $S_{n+1}$, or the expected demand estimated by the supplier earlier, as shown in Eq. (2) and (4),
\begin{equation}\label{atcF}
\ ATC = \frac{F_{c}}{S_{n+1}}+v-vS_{n+1}+(S_{n+1})^2.
\end{equation}

We assume that the variable cost, $v$, and the fix cost, $F_{c}$, are positive constants. The final step in this process is to add profits over the average total cost of the good, using the gross margin operator shown in Eq. (3). When $M$ increases, the price function moves upwards, what leads to higher prices and when it decreases the price moves downwards what leads to cheaper products as shown in Fig.~\ref{ATC}.\

{\bf Assumption 2}\\
{\em The main goal of the supplier is to sell all the produced goods.}

For simplicity, we assume that the supplier cannot keep goods as inventories from one period to the next and also he does not maximize his profits. This model does not take into account the financial constraints of the production process, and we assume that the supplier has money to produce or to buy at any point in time. The main focus of the model is to show how the supplier tries to match his expectations about the demand with the real demand in the market and how this process alters the price. So the question is, how the supplier knows if he had a successful sales campaign. In this case, for him, we consider that a successful sales campaign means that all the goods were sold. This is exactly the market equilibrium assumption except that in our model, is just a temporal state of the system and not a constant reality of the market. The supplier quantifies his success after each period using a very simple model - he divides the quantity demanded at time $n$ by the quantity supplied at time $n$ as shown in Eq. (5). We call it the {\em signal of success} (S),
\begin{equation}\label{Signal}
\ S = \frac{D_{n}}{S_{n}}.
\end{equation}

According to the signal of success, the supplier decides how many goods to produce and supply in the next period of time. From the mathematical point of view, it is important to notice that the supplier reacts to the signal of success and not implicitly to the quantities demanded and supplied. This simple idea helps us to model the market assuming no inventories and inequalities between demand and supply. The signal of success can be divided in four subsets of outcomes, each one with its corresponding economic meaning. We assume that all outcomes are in the positive domain.\\
1. When  $\frac{D_{n}}{S_{n}}=0$, there is no demand, or even worst, there is no market. In this case the supplier will not produce anything for the next period due to the scarcity of demand.\\
2. When $0<\frac{D_{n}}{S_{n}}<1$, the quantity demanded is smaller than the quantity supplied at the given level of price. The supplier produced more goods than what the market could possibly absorb. From the economical point of view, the supplier will probably affront economic losses and also gain negative expectations about the future state of the market.\\
3. When  $\frac{D_{n}}{S_{n}}=1$, the quantity demanded is exactly equal to the quantity supplied. This means that he had a successful sales campaign, exactly as we defined earlier. In general, suppliers aspire to find themselves in this situation. This is a natural equilibrium point of the system as we will show in the following sections.\\
4. When  $\frac{D_{n}}{S_{n}}>1$, the quantity demanded is larger than the quantity supplied. This is a stock-rupture situation. Although the supplier sold all the goods he produced, and this condition meets Assumption 2, losing the possibility to sell even more goods and earn extra revenue, is an unsatisfactory situation for him. Imagine costumers entering through the shop door with money bills in their hands asking for some product that is out of stock. Although he has lost some extra revenue, he gains positive expectations about the future.

The model works as follow, in the first step the supplier supplies some quantity of goods to the market to get some feeling about the demand (seed). Then he observes the quantity of goods that were demanded at the price that he fixed. According to this quantity the supplier decides how many goods to produce or buy for the next period using a simple model that quantify the success of his sales campaign. We called it the signal of success, and it is a simple division between the demanded and supplied quantities at time $n$. After computing the signal of success the supplier uses it to estimate the expected demand in the next period. The second step is the pricing process. The supplier uses his ATC function to compute the goods average total cost. After obtaining the cost per unit, he adds some profits over the cost using the gross margin operator. Finally, he introduces the goods with their new price into the market. He waits some time until he sees how many goods have been sold and then he repeats all the process again.

\section{Two types of suppliers and their behaviors}

\begin{figure}
\begin{center}
\includegraphics[width=0.8 \textwidth]{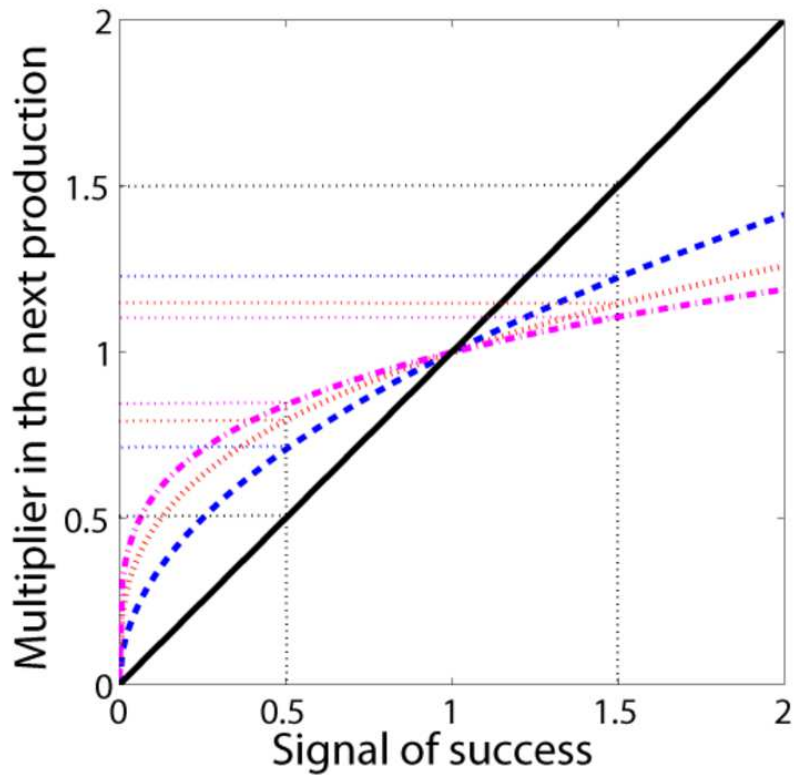}
\caption{\textbf{Behaviors of suppliers in term of the signal of success.} The relationship between the $nth$ root of the signal of success with the multiplier in the next production is shown in the figure above. The solid black curve represents the linear case or the naive supplier, $m=1$. The blue dash line is the square root $m=2$ of the signal of success. The red dot line is the cubic root $m=3$ of the signal of success and the magenta dash-dot line is the $4th$ root of the signal of success. We have plotted the horizontal dot lines, to help the reader see the multiplier of production
 in each case, when the signal of success is $0.5$ and $1.5$. }\label{funciones}
\end{center}
\end{figure}

In this Section we will describe two types of suppliers. Both of them share the function that describes the relationship between the signal of success and the multiplier of production for the next period of time, that is, how the amount of goods produced or bought in the present period of time for the coming sales campaign, is affected by the signal of success. In Fig.~\ref{funciones} we show this relationship. For the sake of simplicity we have used two very simple suppliers that can be modeled analytically. But in the model, more complex suppliers could be introduced.\

{\bf The naive supplier}\\
The simplest assumption of all is that the supplier makes the decision of how many goods to supply in the next period, using the signal of success and the amount of goods he supplied in the previous period as a bench mark. The supplier uses a very simple model to compute the expected demand, that works as follows. He multiplies the signal of success with the quantity supplied in the previous period as shown in Eq. (6),
\begin{equation}\label{SignalNaive}
\ D^{Exp}_{n+1} = (\frac{D_{n}}{S_{n}}) \times S_{n} = D_{n}.
\end{equation}

The logic behind this model is that the supplier expects the demand to behave in the next sales season, exactly the same as it behaved in the previous period. This forecasting method is the same as the moving average method with exponential smoothing coefficient of $\alpha = 1$, putting all the weight of the forecast on the most recent information \cite{smothing}. There is a linear relationship between the signal of success and the multiplier for the next production as shown in Fig.~\ref{funciones}. The supplier is going to produce exactly the same quantity that was demanded in the previous period. For this reason we have called naive, to this supplier. The model takes the following form
\begin{eqnarray}\label{NaiveSupplier}
& D_{n+1} = a - bP_{n+1},\\
& S_{n+1} = D_{n},\label{SNaive}\\
& P_{n+1} =\frac{1}{{1-M}}\cdot(\frac{F_{c}}{S_{n+1}}+v-vS_{n+1}+(S_{n+1})^2). \label{PNaive}
\end{eqnarray}

Simplifying this system of equations, we get the following one dimensional maps for the demand and the price,
\begin{equation}\label{recursiveDNaive}
\ D_{n+1}(1-M) = a(1-M)-b(\frac{F_{c}}{D_{n}}+v-vD_{n}+(D_{n})^2),
\end{equation}

\begin{equation}\label{recursivePNaive}
\ P_{n+1}=\frac{1}{{1-M}}\cdot(\frac{F_{c}}{a-b(P_{n})}+v-v(a-b(P_{n})+(a-b(P_{n}))^2).
\end{equation}\

{\bf The cautious and optimistic supplier}\\
This type of supplier is in fact a family of infinite number of suppliers, each one with a different sensitivity to the signal of success. This supplier instead of merely using the signal of success as it is, prefers to transform it to be able to improve the prediction of the demand in the next period. He uses a very simple but powerful model. He finds the $nth$ root of the signal of success where $m$ defines his cautiousness and optimism as we will see next. The supplier multiplies the $nth$ root of the signal of success with the quantity supplied in the previous period that serves him as bench mark. We can see this model in Eq. (12),
\begin{equation}\label{SignalCOsupplier}
D^{Exp}_{n+1} = \sqrt[m]{(\frac{D_{n}}{S_{n}})} \times S_{n},
\end{equation}
where $m > 0$. From Fig.~\ref{funciones} we can see that when $m$ increases the supplier becomes less optimistic and more cautious about the future state of the market, when the signal of success is greater than one. But he becomes less cautious and more optimistic when the signal of success is between zero and one. This behavior remaind loss aversion \cite{loss}, where the suppliers reference point, is when the signal of success is equal to one. As the reader might guess the naive supplier is just a particular case in this model and it arises when $m=1$.

So $m$ determines the producer's sensitivity to the market states or to the signal of success perceived. In general, all of them behave in the same manner.
When $\frac{D_{n}}{S_{n}}=0$, and  $\frac{D_{n}}{S_{n}}=1$, there is no change in their behaviors, they expect the demand to be $0$ and $D_{n}$ respectively as we saw in the naive supplier case. The interesting behavior occurs when $0<\frac{D_{n}}{S_{n}}<1$, and when $\frac{D_{n}}{S_{n}}>1$. In the first subset of outcomes the supplier perceives lower demand in proportion to the quantity supplied at time $n$. Because of that, he will produce fewer goods than before. His optimism will drive him to produce a little bit more goods compared to what the naive producer would had produced in the same situation. As his $m$ increases the supplier becomes more and more optimistic and he will produce more goods. On the other hand when $\frac{D_{n}}{S_{n}}>1$, the supplier perceives high demand in proportion to the quantity supplied at time $n$. Therefore, he will produce more goods than before. However, his cautiousness will play an important role. He will produce fewer goods compared to what a naive producer had produced in the same situation. As his $m$ increases he is considered to be more cautious and he will produce less goods.
We can write down this model as follow,
\begin{equation}
\ D_{n+1} = a - bP_{n+1},
\end{equation}
\begin{equation}
\ S_{n+1} =\sqrt[m]{(\frac{D_{n}}{S_{n}})} \times S_{n},
\end{equation}
\begin{equation}
\ P_{n+1} =\frac{1}{{1-M}}\cdot(\frac{F_{c}}{S_{n+1}}+v-vS_{n+1}+(S_{n+1})^2). \label{PCOsupplier}
\end{equation}

Simplifying this system of equations we obtain the following two dimensional map for the demand and the supply,
\begin{equation}
\ D_{n+1}(1-M) = a(1-M) - b(\frac{F_{c}}{S_{n+1}}+v-vS_{n+1}+(S_{n+1})^2).
\end{equation}
\begin{equation}
\ S_{n+1} = \sqrt[m]{(\frac{D_{n}}{S_{n}})} \times S_{n}.
\end{equation}

Here the producer needs two seeds to calculate the expected demand, $D_{0}$ and, $S_{0}$. Notice that this two dimensional map can be reduce into a one dimensional map in terms of supply as shown in Eq. (18).

\begin{equation}
\ S_{n+1} = \sqrt[2]{\frac{1}{S_{n}}(\frac{1}{1-M}(a-b(\frac{F_{c}}{S_{n}}+v-vS_{n}+S_{n}^2)))} \times S_{n}.
\end{equation}

\section{Methodology}

We have studied only two variations of the model. Equation (19), shows the naive supplier when the parameters are fixed as: $a =10$, $b = 0.09$, $v = 4$, $Fc = 10$ and $M = 0.5$.
\begin{equation}\label{recursiveDNaive}
\ D_{n+1}(0.5) = 10 - 0.09(\frac{10}{D_{n}}+4-4D_{n}+(D_{n})^2).
\end{equation}

Equations (20) and (21) represent the cautious and optimistic supplier when the parameters are fixed as:      $a = 30$, $b = 0.125$, $v = 6$, $Fc = 30$, $M = 0.5$ and $m = 2$.

\begin{equation}
\ D_{n+1}(0.5) = 15 - 0.125(\frac{30}{S_{n+1}}+6-6S_{n+1}+(S_{n+1})^2),
\end{equation}

\begin{equation}
\ S_{n+1} = \sqrt[2]{(\frac{D_{n}}{S_{n}})} \times S_{n}.
\end{equation}

We have used the following one dimensional map to compute the Lyapunov exponents spectrum of the cautious and optimistic supplier,

\begin{equation}
\ S_{n+1} = \sqrt[2]{\frac{1}{S_{n}}(2(30-b(\frac{30}{S_{n}}+6-6S_{n}+S_{n}^2)))} \times S_{n}.
\end{equation}

We have studied the dynamics of both models using three tests. First, we have computed the time series of both models to observe the dynamics by applying a recursive algorithm. We have changed the parameters $b$ and $M$ to see how the dynamics of the time series changes. We have chosen to show only the chaotic time series because we want to prove the existence of chaos in the model. Secondly, we have plotted the bifurcation diagrams of the quantity demanded against the parameter $b$ in both cases. At each value of $b$, we have iterated the functions until they reached the equilibrium points using a recursive algorithm. Then, we have plotted the values of $D_{n+1}$ corresponding to the specific value of $b$ on the same plot. We have done the same with the parameter $M$ in the naive supplier case, to show the dynamics when the margin is changed. Lastly, we have computed the Lyapunov exponents spectrum of both systems.

\section{Global dynamics and results}

\begin{figure}
\begin{center}
\includegraphics[width=0.8 \textwidth]{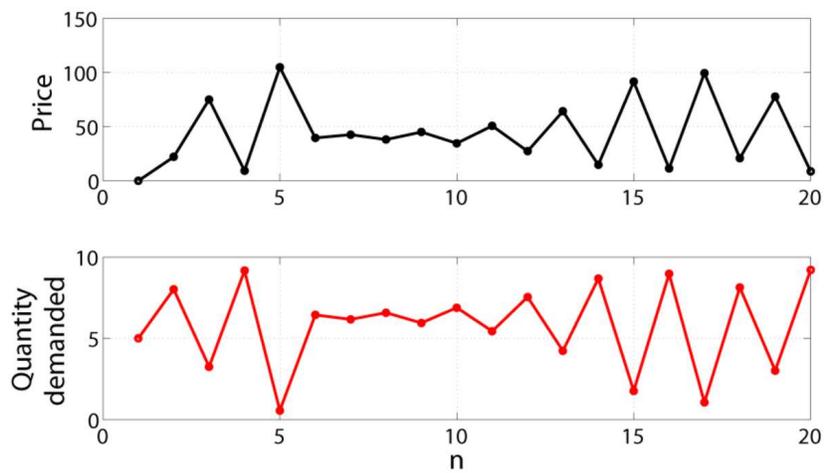}
\caption{\textbf{The price and demand time series that correspond to the naive supplier during the first $20$ periods of trade.} The two time series that are shown in the figure above were plotted iterating Eq. (18) and (10). The black line corresponds to the price, and the red line corresponds to the demanded quantity in the first 20 periods of trade. Despite the fact that the price and the demand are discrete quantities, it is more easy to follow their evolution plotting them as continuous curves. But, note that the lines between the dots are
meaningless.}\label{NaiveTS}
\end{center}
\end{figure}

{\bf The naive supplier}\\
In order to understand the relationship between the price and the quantity demanded, we have plotted the first $20$ periods of trade as shown in Fig.~\ref{NaiveTS}. We clearly see the price and the quantity demanded behave exactly how we expected. High prices are responded with low demand and low prices are responded with high demand. However, the plots show an irregular behavior in both cases. The economical meaning of this behavior is that the supplier and the customers have not agreed on the quantity and the price during the trade. In other words, their interactions were not translated into market equilibrium. Furthermore, it seems that this market is not efficient. But there is a small window between time steps $6$ to $10$, in which the trajectories of the price and the quantity demanded are almost flat or almost in equilibrium. However, after two time steps this behavior changes abruptly into high amplitude fluctuations. We would expect that real world markets of ordinary goods, to behave dynamically and not to fall into the frozen state that standard models predict. We did not obtain this behavior by an accident; we have chosen the parameter values precisely to get this behavior. Next, we will show that more dynamical behaviors are possible computing the bifurcation diagram.

\begin{figure}
\begin{center}
\includegraphics[width=0.8 \textwidth]{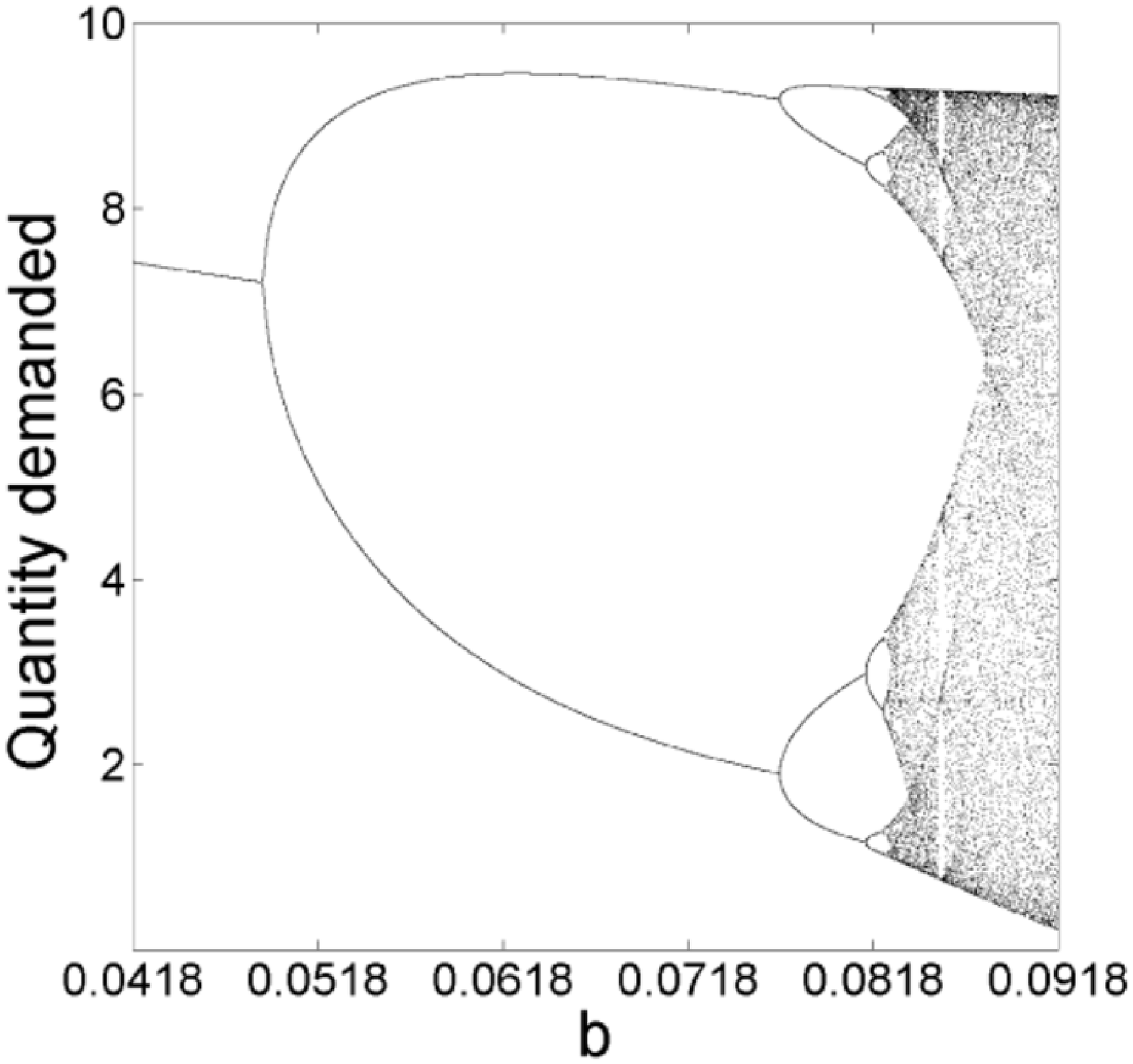}
\caption{\textbf{The bifurcation diagram of the quantity demanded, $D_{n+1}$, against the parameter $b$.}
We have divided the interval $(0.0418, 0.0918)$ of the parameter $b$ into $10,000$ values. Then, we have set each value of the parameter $b$ in Eq. (10) and we have iterated the equation $3,000$ times until it settles down in the corresponding fixed points. Finally, we have plotted those fixed points against the value of the parameter $b$ to obtain this bifurcation diagram.}
\label{bifurNaiveProducer}
\end{center}
\end{figure}

For given values of the parameters $b$, and $M$, we can compute the fixed points of the Eq. (11). If we allow the parameter $b$ to vary between $0$ and $0.0918$, we can establish the equilibrium points for $D_{n+1}$, by plotting the bifurcation diagram of $D_{n+1}$ against $b$ as shown in Fig.~\ref{bifurNaiveProducer}.

\begin{figure}
\begin{center}
\includegraphics[width=0.8 \textwidth]{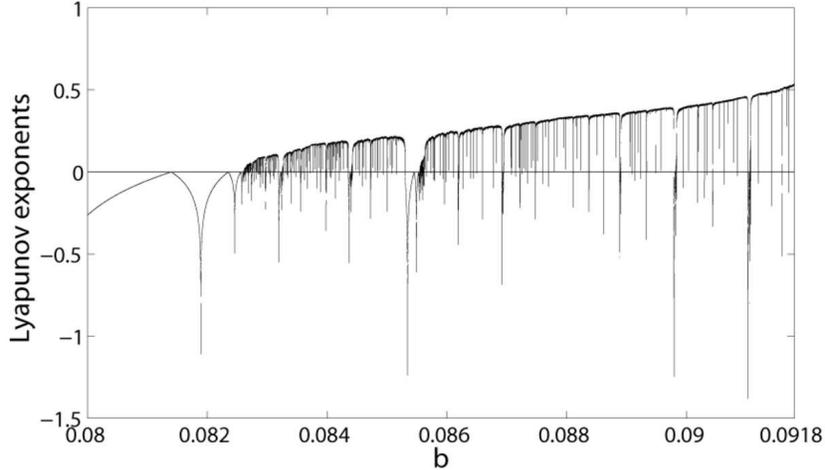}
\caption{\textbf{The Lyapunov exponent's spectrum corresponding to the naive supplier when parameter $b$ is varied.} We have taken the interval $(0.08,0.092)$ of the parameter $b$ and we have computed the Lyapunov exponent of $100,000$ points within this interval. Finally, we have plotted the corresponding exponent against its corresponding value of the parameter $b$ to obtain the spectrum. The exponent is positive in a wide range of parameter $b$ values, what proves the chaotic behavior of the system.
}\label{Lyapunov}
\end{center}
\end{figure}

\begin{figure}
\begin{center}
\includegraphics[width=0.8 \textwidth]{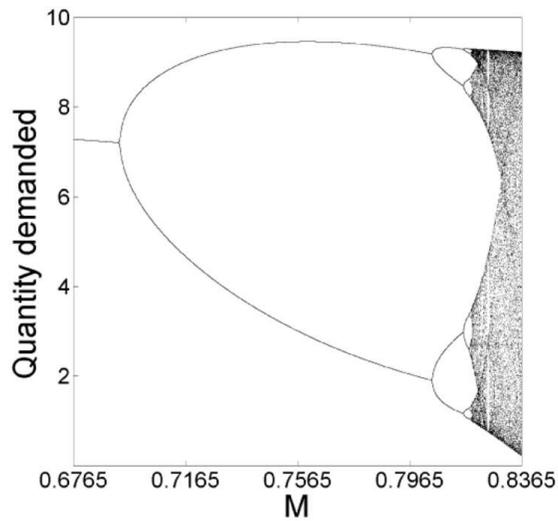}
\caption{\textbf{The bifurcation diagram of the quantity demanded, $D_{n+1}$, against the parameter $M$.}
We have divided the interval $(0.6765, 0.8365)$ of the parameter $M$ into $20,000$ values. Then, we have set each value of parameter $M$ in Eq. (10) and we have iterated the equation $3,000$ times until it settles down in the corresponding fixed points. Finally, we have plotted those fixed points against the value of parameter $M$ to obtain this bifurcation diagram. Notice that when the gross margin is between $0$ and $0.6765$ the system is in equilibrium. This is a huge range of gross margin values. In contrast, only a small part of the gross margin interval causes the demand to behave chaotically. It is not a surprise that this small part corresponds to high margins. }\label{bifurNaiveProducerMargin}
\end{center}
\end{figure}

The period-doubling route to chaos \cite{Yorke, Feingenbaum} is obvious looking at Fig.~\ref{bifurNaiveProducer}. We have found period $6$ and period $10$ cycles when $b = 0.8531$ and $b = 0.0843999995$, respectively. We clearly see the huge range of demand dynamics when we are varying the parameter $b$. We will explain why this outcome is meaningful in terms of demand theory in the next Section. We obtain a similar bifurcation diagram when we vary $M$ against $D_{n+1}$. Figure ~\ref{bifurNaiveProducerMargin} shows how the quantity demanded is affected by the gross margin, when it is changed. Notice that in Fig.~\ref{bifurNaiveProducerMargin}, $b = 0.03$. One can check in Fig.~\ref{bifurNaiveProducer} that at this value of the parameter $b$, the system should be in equilibrium. Incrementing the gross margin in order to obtain more profits leads to a destabilization of the whole system. The model suggests that the supplier greed has limits. This is the proof that the supplier has influence on the global dynamics of market. We have also computed the Lyapunov exponent spectrum to prove the existence of chaos as shown in Fig.~\ref{Lyapunov}.

\begin{figure}
\begin{center}
\includegraphics[width=0.8 \textwidth]{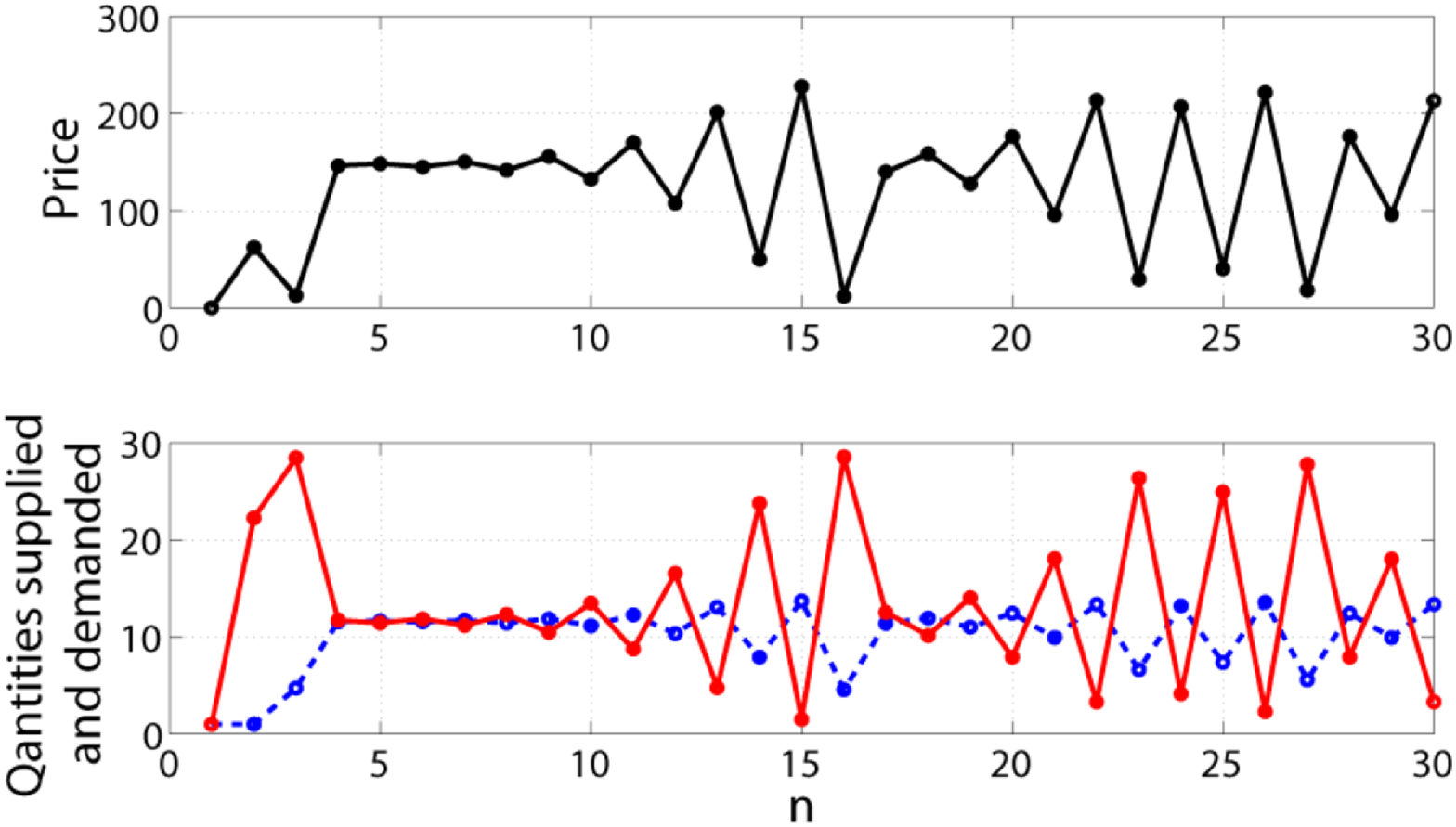}
\caption{\textbf{Time series of the cautious and optimistic supplier in the first $30$ periods of trade.} At the bottom we have plotted the demand $D$ as a solid red line against the supply $S$ as a dash blue line. Above in black, we have plotted the price trajectory in this trade scenario. This figure shows the dynamic behavior of the quantities supplied and demanded, and the price. The price is moving exactly as we would expect. There are periods where the price does not change much, so we can say the market is almost in equilibrium. And there are periods where the price changes dramatically, what corresponds to the nonequilibrium state of the market.
}\label{optimisticTS}
\end{center}
\end{figure}

{\bf The cautious and optimistic supplier}\\
We start again with the time series shown in Fig.~\ref{optimisticTS}.

\begin{figure}
\begin{center}
\includegraphics[width=0.8 \textwidth]{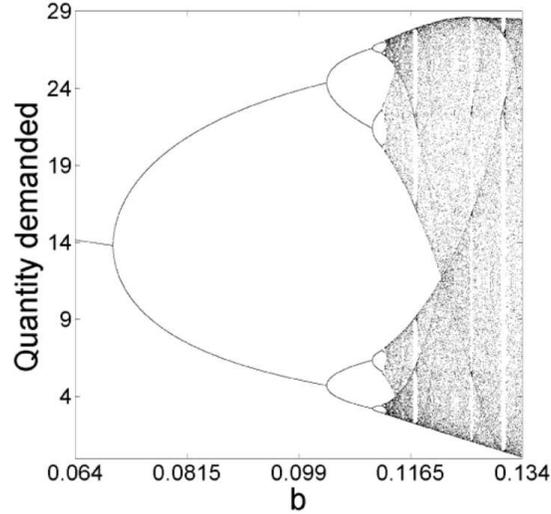}
\caption{\textbf{The bifurcation diagram of the quantity demanded, $D_{n+1}$, against the parameter $b$.}
We have divided the interval $(0.064, 0.134)$ of the parameter $b$ into $10,000$ values. Then, we have set each value of the parameter $b$ in Eq. (16) and we have iterated the equation $3,000$ times until it settles down in the corresponding fixed points. Finally, we have plotted those fixed points against the value of the parameter $b$ to obtain this bifurcation diagram. }
\label{bifOptimisticProducer}
\end{center}
\end{figure}

\begin{figure}
\begin{center}
\includegraphics[width=0.8 \textwidth]{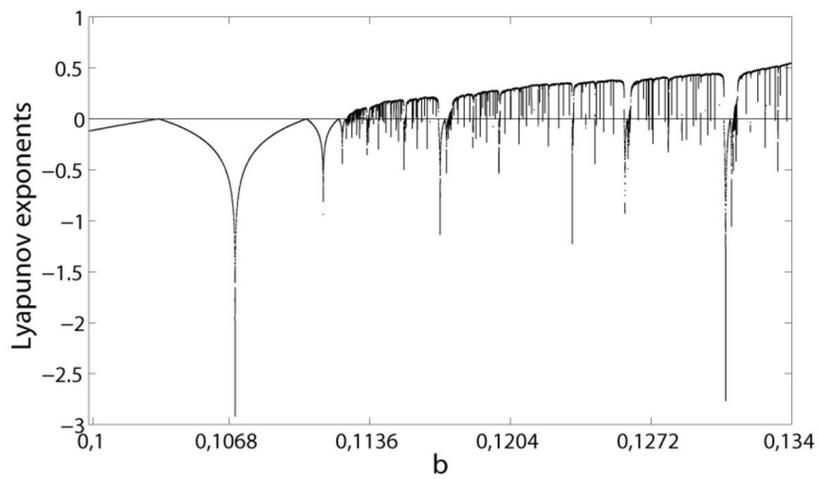}
\caption{\textbf{The Lyapunov exponent's spectrum corresponding to the cautious and optimistic supplier when parameter $b$ is varied.} We have taken the interval $(0.1,0.134)$ of the parameter $b$ and we have computed the Lyapunov exponent of $80,000$ points within this interval. Finally, we have plotted the corresponding exponent against its corresponding value of parameter $b$ to obtain the spectrum. The exponent is positive in a wide range of parameter $b$ values, what proves the chaotic behavior of the system.
}\label{OptimisticLyapunov}
\end{center}
\end{figure}

 It is possible to verify how high prices are responded with low demand and vice versa. We can see periods where the demanded and supplied quantities are almost the same. In these periods the system is almost at equilibrium so the price is stable. But after some time the system goes out of equilibrium and periodic-cycles and chaotic behavior arise. We have plotted the bifurcation diagram of $D_{n+1}$ against $b$ to illustrate some more possible behaviors as shown in Fig.~\ref{bifOptimisticProducer}. A period 3 cycle occurs when $b = 0.1308$. This observation implies chaos \cite{Period three}. We can clearly see that the period doubling route to chaos from Fig.~\ref{bifOptimisticProducer} as well. Furthermore, we have computed the Lyapunov exponent spectrum to prove the existence of chaos as shown in Fig.~\ref{OptimisticLyapunov}.

\section{The final bifurcation means market collapse}

In this Section, we will expand the economical assumptions of the model to emphasize the idea, that a final bifurcation can be a good description of a market collapse. We have chosen the naive supplier as a case study. But the reasoning and the methodology that we have used to demonstrate this claim, is generic, and can be applied to all types of suppliers.

When the parameters are fixed in Eq. (10), and (11) as: $D_{1} = 1$, $S_{1} = 1$, $a = 10$, $b = 0.095$, $v = 2$, $Fc = 20$ and $M = 0.5$, we get the following maps for the demand, the supply and the price :
\begin{equation}\label{recursiveDNaive}
\ D_{n+1}(0.5) = 10 - 0.095(\frac{20}{D_{n}}+2-2D_{n}+(D_{n})^2).
\end{equation}

\begin{equation}
\ S_{n+1} = \sqrt[2]{\frac{1}{S_{n}}(2(10-0.095(\frac{20}{S_{n}}+2-2S_{n}+S_{n}^2)))} \times S_{n}.
\end{equation}

\begin{equation}\label{recursivePNaive}
\ P_{n+1}=\frac{\frac{20}{10-0.095(P_{n})}+2-2(10-0.095(P_{n}))+(10-0.095(P_{n}))^2}{1-0.5}.
\end{equation}\

\begin{figure}
\begin{center}
\includegraphics[width=0.8 \textwidth]{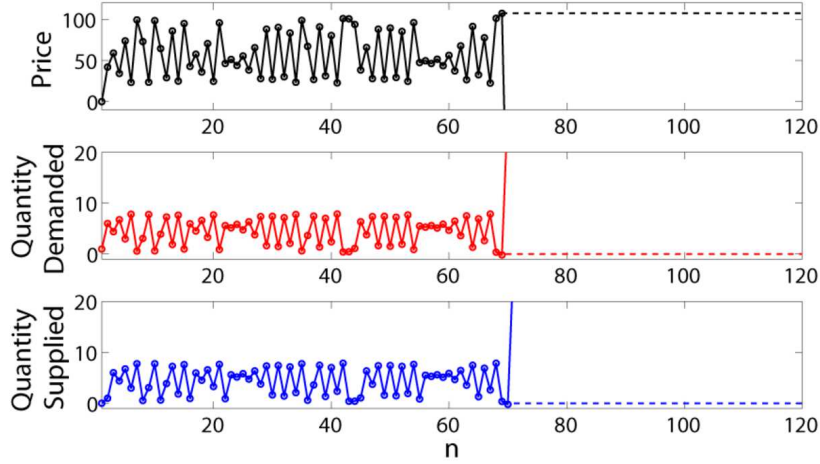}
\caption{\textbf{Time series of the quantities demanded and supplied before and after bounding the system.}  The solid line represent the time series of the price, the demand and the supply, simply by iterating the maps fixing the parameters as: $D_{1} = 1$, $a = 10$, $b = 0.095$, $v = 2$, $Fc = 20$ and $M = 0.5$. The time series behaves chaotically until time step $69$ where a very big fluctuation occurs. The price becomes negative so the quantities demanded and supplied increase dramatically. The dash line represents the same system as before but now bounded. The time series can not be negative so that, when some critical value is crossed the system simply goes to zero, as in the case of the quantity demanded and supplied shown in the figure above.   }\label{Naive1}
\end{center}
\end{figure}

Analyzing the time series produced by these maps we find a transient chaotic behavior as shown in Fig.~\ref{Naive1}. The trajectories of the quantities demanded, the quantity supplied and the price are completely chaotic until time step $69$, where suddenly they explode. By explode we mean the system starts to fluctuate without control giving rise to quantities that are unscaled to the system or even infinitely large. We are not familiar with the complicated concepts of negative infinite price or infinite demand and supply. Therefore, to get a better economical understanding of this situation we need to extend our assumptions about the model.

We will first, focus on the demand side of the system. The meaning of parameter $a$ in Eq. (1) is that when the good is freely available (its price is zero) in the market, the maximum amount of goods that can be demanded is the value of the parameter $a$. This is an accomplished fact, and it is the upper bound of the quantity of goods that can be demanded in this market, assuming the system lies in the positive domain. When we allowed the price to take negative values, the amount of goods demanded was much higher from the value of the parameter $a$. In this scenario the supplier must pay the consumer to create the demand. We will assume that the supplier does not make strategic decisions thinking on long time horizons. So, when the price is negative he just lose the incentives to supply. Equation (26) integrates this new behavior into the model,
\begin{equation}
D_{n+1} = \left\lbrace
\begin{array}{ll}
\textup 0 & {if } \; (P_{n+1} \times b > a),\\
\textup a - b \times P_{n+1} & {if } \; (P_{n+1} \times b \leq a),
\end{array}
\right.
\end{equation}

Following the same reasoning as in the demand case, we extend our assumptions on the supply side of the system. The second assumption of the model is that the supplier always tries to sell exactly the amount of goods he produced or bought. If he expects zero or negative demand we can assume the supplier will not produce anything for the next period of time. He will probably get out of the market in this situation. The supplier computes the expected demand before going into production, so if he sees that the expected demand is zero or negative he stops immediately the process. We can describe mathematically this behavior using Eq. (27).
\begin{equation}
S_{n+1} = \left\lbrace
\begin{array}{ll}
 & \frac{1}{{1-M}}\cdot( \frac{F_{c}}{S_{n+1}}+v-vS_{n+1}+(S_{n+1})^2) \; \textup{if } D_{n+1} > 0,\\
 & stop \; \textup {if } D_{n+1}, \leq 0
\end{array}
\right.
\end{equation}

When the trajectories arrive to the final bifurcation the market stops to exist immediately. The reader can see in Fig.~\ref{Naive1}, how after the final bifurcation the price stays at some high level where the quantities supplied and demanded go to zero. Note that if the demand crosses some critical value (small value), the system enter into a loop of destruction, because of the growing cost of production of diminishing quantities. We would expect similar dynamics in a situation of market collapse.

In real economies we find two interesting properties that can be also observed in this model. The first one is the prediction problem, in which the collapse is impossible to forecast beforehand. Secondly, the global complexity of the market emerges from simple nonlinear interactions between the economical agents.

\section{The influence of the price elasticity of demand (PED) on the global dynamics}

We have modeled the demand as a monotonic function. Nevertheless, the slope of the demand curve, parameter $b$, has a huge effect on the dynamics of the system as we saw in the previous sections. To capture this idea we can compute the {\em price elasticity of demand} (PED), which measures the quantity demanded sensitivity to the price and it is given by the following ratio:
\begin{equation}
\ PED = \frac{\% \text{change in Quantity demanded}}{\% \text {change in Price}}.\label{PED}
\end{equation}

\begin{figure}
\begin{center}
\includegraphics[width=0.8 \textwidth]{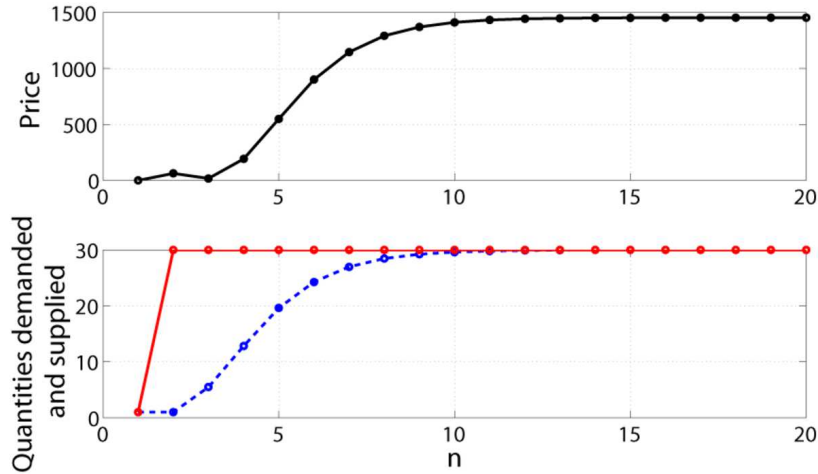}
\caption{\textbf{Dynamics of supply when there is a perfectly elastic demand curve.} Time series of the first $20$ periods of trade in the cautious and optimistic producer case when $b = 0$. At the bottom we plotted the demand $D$ in red against the supply $S$ in blue. Above we plotted the price trajectory of this trade scenario.}
\label{OptimisicTSbzero}
\end{center}
\end{figure}

In general, goods which are elastic tend to have many substitutes, they must be bought frequently and they assume to be traded in a very competitive market. In this model we have assumed all above. We have done this by modeling the market as an ordinary good market that obeys the demand law. When we vary the parameter $b$, we change the price elasticity of demand. For example, when $b = 0$, we encounter a perfectly elastic demand curve. One can imagine the demand curve as an horizontal line. At this certain price the demand is infinite, so any amount of goods is quickly consumed. In Fig.~\ref{OptimisicTSbzero} we clearly see how the quantity supplied in blue is rapidly sticking to the quantity demanded in red until all the demand is fulfilled. Due to the excess demand the price is going up until it reaches the market equilibrium price. This process is not instantaneous as can be checked. Even though we have assumed a perfectly elastic demand, the supplier does not know it. It takes him about $13$ periods of trade to supply all the goods demanded by the market. This is a good example of the adjustment dynamics that underlies the market equilibrium assumption.

But the really remarkable result is that a very small change in the PED can change completely the system dynamics. Figure ~\ref{bifOptimisticProducer} describes how the global dynamics of the system changes as we increase the value of the parameter $b$ inside a very small subset. When $0 < b < 0.134$, we observe equilibrium points, cycles and chaotic trajectories, but when $b > 0.134$, the system explodes. We have showed in the previous Section that the economical meaning of this exploding dynamics is a market collapse.

This behavior is not special only for $b$, when the value of $M$ and $a$ are varied, we encounter the same dynamics, but we assume that the gross margin value is controlled or partially controlled by the supplier. Therefore, theoretically the supplier can avoid erratic trajectories or crash scenarios manipulating this variable. We have focused on the price elasticity of demand because it cannot be influenced by the supplier but it is directly related to the price. Exactly like in a real world, the small supplier tries to adjust its production to the demand, and not the demand to the production. Because trying to influence the demand is highly expensive and only big companies with more resources can afford it.

\section{Conclusions}

We have introduced the supply based on demand model studying two types of suppliers, the naive supplier and the cautious and optimistic supplier. In both cases we have found that the model is capable of reproduce a large large variety of dynamics such as equilibrium, cycles, chaos, and even catastrophic dynamics under simple and reasonable economic assumptions. We have emphasized the idea that the final bifurcation can be a good description of a market collapse by adding some new assumptions to the model. We have shown the important role that the price elasticity of demand plays on the global dynamics of the market. One important result is that very small changes in the price elasticity of demand leads to very different global dynamics assuming a monotonic demand function. We have also demonstrated the huge influence of the gross margin, $M$, on the market dynamics.

\section*{Acknowledgments}

This work was supported by the Spanish
Ministry of Economy and Competitiveness under project number
FIS2013-40653-P and FIS2016-76883-P.


\bigskip




\begin{thebibliography}{1}


\bibitem[Graves, 2011]{Uncertenty and Production Planning}
Graves SC.
\newblock {{U}ncertainty and {P}roduction {P}lanning}.
\newblock Planning  Production and Inventories in the Extended Enterprise, A State of the Art Handbook, Kempf K.G., Keskinocak P., Uzsoy R. (Eds.), Series: International Series in Operations Research \&
Management Science 2011;151:83--102.

\bibitem[Akerlof, 1970]{lemons}
Akerlof GA.
\newblock {{T}he {M}arket for ``{L}emons'': {Q}uality {U}ncertainty and the {M}arket {M}echanism}.
\newblock Q. J. Econ. 1970;84:488--500.

\bibitem[Baron, 1971]{Baron}
Baron DP.
\newblock {{D}emand {U}ncertainty in {I}mperfect {C}ompetition}.
\newblock Int. Econ. Rev. 1971;12:196--208.

\bibitem[Dana, 2001]{Dana}
Dana JD.
\newblock {{M}onopoly price dispersion under demand uncertainty}.
\newblock Int. Econ. Rev. 2001;42:649--670.

\bibitem[Milton \& Raviv, 1981]{Raviv}
Milton H, Raviv A.
\newblock {{A} {T}heory of {M}onopoly {P}ricing {S}chemes with {D}emand {U}ncertainty}.
\newblock Am. Econ. Rev. 1981;71:347--365.

\bibitem[Arthur, 2006]{Brian Arthur}
Arthur BW.
\newblock {{A}gent-{B}ased {M}odeling and {O}ut-{O}f-{E}quilibrium {E}conomics}.
\newblock In: L. Tesfatsion and K.L. (Eds.), Handbook of Computational
Economics, Amsterdam: Elsevier Science 2006;2:1551--1564.

\bibitem[Farmer \& Foley, 2009]{Farmer}
Farmer JD, Foley D.
\newblock {{T}he economy needs agent-based modeling}.
\newblock Nature 2009;460:685--686.

\bibitem[Tesfatsion, 2006]{Tesfatsion}
Tesfatsion L.
\newblock {{A}gent-based computational economics: a constructive approach to economic theory}
\newblock In: L. Tesfatsion and K.L. Judd (Eds.), Handbook of Computational
Economics, Amsterdam: Elsevier Science 2006;2:831--880.

\bibitem[Edge et al., 2008]{dsge}
Edge RM, Kiley MT, Laforte JP.
\newblock {{N}atural {R}ate {M}easures in an {E}stimated DSGE {M}odel of the U.S. {E}conomy}.
\newblock J. Econ. Dyn. Control 2008;32:2512--35.

\bibitem[Gertler et al., 2008]{dsgeW}
Gertler M, Sala L, Trigari A.
\newblock {{A}n {E}stimated {M}onetary DSGE {M}odel with {U}nemployment and {S}taggered {N}ominal {W}age {B}argaining}.
\newblock J. Money Credit Banking 2008;40:1713--64.

\bibitem[Sbordone et al., 2010]{introDsge}
Sbordone AM, Tambalotti A, Krishna R, Kieran JW.
\newblock {{P}olicy {A}nalysis {U}sing DSGE {M}odels: {A}n {I}ntroduction}.
\newblock Economic Policy Review 2010;1:23--43.

\bibitem[Chase, 2009]{Chase}
Chase CW.
\newblock {Demand-Driven Forecasting: A Structured Approach to Forecasting Second Edition}.
\newblock Wiley and SAS Business Series; 2009.

\bibitem[Ezekiel, 1938]{Cobweb theorem}
Ezekiel M.
\newblock {{T}he cobweb theorem}.
\newblock Q. J. Econ. 1938;52:255--280.

\bibitem[Hommes, 1994]{Hommes}
Hommes CH.
\newblock {{D}ynamics of cobweb model with adaptive expectations and nonlinear supply and demand}.
\newblock J. Econ. Behav. Organ. 1994;24:315--335.

\bibitem[Nerlove, 1958]{Nerlove}
Nerlove M.
\newblock {{A}daptive expectations and cobweb phenomena}.
\newblock Q. J. Econ. 1958;72:227--240.

\bibitem[Chiarella, 1988]{Chiarella}
Chiarella C.
\newblock {{T}he cobweb model. {I}ts instabilities and the onset of chaos}.
\newblock Econ. Model. 1988;5:377--384.

\bibitem[Artstein, 1938]{Artstein}
Artstein Z.
\newblock {{I}rregular cobweb dynamics}.
\newblock Econ. Lett. 1938;11:15--17.

\bibitem[Holt et al., 1960]{Holt Simon}
Holt CC, Modigliani F, Muth JF, Simon HA.
\newblock {Planning Production, Inventories and Work Force}.
\newblock Englewood Cliffs NJ: Prentice-Hall; 1960.


\bibitem[Gardner, 2006]{smothing}
Gardner JE.
\newblock {{E}xponential smoothing: the state of the art - part II}.
\newblock I. J. Forecasting 2006;22:637--666.

\bibitem[Tversky \& Daniel, 1991]{loss}
Tversky A, Kahneman D.
\newblock {{L}oss {A}version in {R}iskless {C}hoice: {A} {R}eference {D}ependent {M}odel}.
\newblock Q. J. Econ. 1991;107:1039--1061.


\bibitem[Yorke \& Alligood, 1985]{Yorke}
Yorke JA, Alligood KT.
\newblock {{P}eriod doubling cascades of attractors: a prerequisite for horseshoes}.
\newblock Commun. Math. Phys. 1985;101:305--321.

\bibitem[Feigenbaum, 1979]{Feingenbaum}
Feigenbaum MJ.
\newblock {{T}he universal metric properties of nonlinear transformations}.
\newblock J. Stat. Phys. 1979;21:669--706.

\bibitem[Li \& Yorke, 1975]{Period three}
Li TY, Yorke JA.
\newblock {{P}eriod {T}hree {I}mplies {C}haos}.
\newblock Am. Math. Mon. 1975;82:985--992.


\end{thebibliography}
\end{document}